\documentclass[10pt,conference]{IEEEtran}
\IEEEoverridecommandlockouts
\usepackage{cite}
\usepackage{amsmath,amssymb,amsfonts}
\usepackage{algorithmic}
\usepackage{graphicx}
\usepackage{textcomp}
\usepackage{xcolor}
\usepackage{paralist}
\usepackage{subcaption}
\def\BibTeX{{\rm B\kern-.05em{\sc i\kern-.025em b}\kern-.08em
    T\kern-.1667em\lower.7ex\hbox{E}\kern-.125emX}}
\begin{document}

\title{Performance of Commercial Quantum Annealing Solvers for the
Capacitated Vehicle Routing Problem\\

}

\author{\IEEEauthorblockN{1\textsuperscript{st} Salvatore Sinno}
\IEEEauthorblockA{\textit{School of Computing} \\
\textit{Newcastle University}\\
Newcastle, UK \\
S.Sinno2@newcastle.ac.uk}

\and
\IEEEauthorblockN{2\textsuperscript{nd} Thomas Gro{\ss}}
\IEEEauthorblockA{\textit{School of Computing} \\
\textit{Newcastle University}\\
Newcastle, UK \\
}
\and
\IEEEauthorblockN{3\textsuperscript{rd} Alan Mott}
\IEEEauthorblockA{\textit{Enterprise Computing Solutions} \\
\textit{Unisys UK Ltd}\\
Milton Keynes, UK \\
}
\and
\IEEEauthorblockN{4\textsuperscript{th} Arati Sahoo}
\IEEEauthorblockA{\textit{Enterprise Computing Solutions} \\
\textit{Unisys India Pvt.}\\
Bangalore, India \\}
\and
\IEEEauthorblockN{5\textsuperscript{th} Deepak Honnalli}
\IEEEauthorblockA{\textit{Enterprise Computing Solutions} \\
\textit{Unisys India Pvt.}\\
Bangalore, India \\}
\and
\IEEEauthorblockN{6\textsuperscript{th} Shruthi Thuravakkath}
\IEEEauthorblockA{\textit{Enterprise Computing Solutions} \\
\textit{Unisys India Pvt.}\\
Bangalore, India \\}
\and
\IEEEauthorblockN{7\textsuperscript{th} Bhavika Bhalgamiya}
\IEEEauthorblockA{\textit{Enterprise Computing Solutions} \\
\textit{Unisys}\\
Pennsylvania, USA \\
bhavika.bhalgamiya@unisys.com
}
}

\maketitle

\begin{abstract}
Quantum annealing (QA) is a heuristic search algorithm that can run on Adiabatic Quantum Computation (AQC) processors to solve combinatorial optimisation problems. Although theoretical studies and simulations on classic hardware have shown encouraging results, these analyses often assume that the computation occurs in adiabatically closed systems without environmental interference. This is not a realistic assumption for real systems; therefore, without extensive empirical measurements on real quantum platforms, theory-based predictions, simulations on classical hardware or limited tests do not accurately assess the current commercial capabilities.
This study has assessed the quality of the solution provided by a commercial quantum annealing platform compared to known solutions for the Capacitated Vehicle Routing Problem (CVRP). The study has conducted extensive analysis over more than 30 hours of access to QA commercial platforms to investigate how the size of the problem and its complexity impact the solution accuracy and the time used to find a solution. Our results have found that the absolute error is between 0.12 and 0.55, and the quantum processor unit (QPU) time is between 30 and 46 $\mu$s. Our results show that as the constraint density increases, the quality of the solution degrades. Therefore, more than the problem size, the model complexity plays a critical role, and practical applications should select formulations that minimise the constraint density.
\end{abstract}

\begin{IEEEkeywords}
quantum annealing,  CVRP, quantum optimisation, D-Wave, CQM, QUBO
\end{IEEEkeywords}

\section{Introduction}
Quantum annealing (QA)  \cite{19} is a heuristic search algorithm that can run on an Adiabatic Quantum Computation (AQC) platform using quantum properties such as tunnelling, entanglement and superposition to solve combinatorial optimisation problems. Like Simulated Annealing (SA), QA uses a parameter called tunnelling coefficient (\(\Gamma\)) to control the transversability of the solution landscape and the probability of taking an uphill step at each iteration.
QA is emerging as a promising generic approach for tackling complex optimisation problems as recent advancements have made this technology commercially available.
Although theoretical studies and simulations on classic hardware (e.g. Path Integral Monte Carlo (PMIC)) have shown encouraging results, these analyses often assume that the computation occurs in adiabatically closed systems with no environmental interference. With this assumption, the algorithm is probabilistic and the Quantum Adiabatic Theorem may bind the trade-off between computation time and the probability that the system found an optimal solution.  However, all real-world quantum computations occur in open systems, vulnerable to environmental noise that reduces the probability of finding a good solution and increases computation time. 
For these reasons, without extensive empirical measurements on real quantum platforms, theory-based predictions, simulations on classical hardware or limited tests do not provide a reliable assessment of the current commercial capabilities.
This study assesses the quality of the solution provided by a commercial quantum annealing platform for the CVRP problem, a well-known logistic combinatorial optimisation problem.
A well-known  limitation of quantum computing is the problem size: it is, therefore, important to investigate how the complexity of the problem (size and constraint density) impacts the quality of the solution.
This study considers many simulations (100 for each instance) over more than 30 hours of access to QA commercial platforms.
The rest of the paper is organised as follows: Section \ref{section:RelatedWork} analyses previous studies using quantum annealing to solve the CVRP problem. Section \ref{Aim} describes the aims of this analysis and the measures used. Section \ref{section:Method} describes the method followed, outlining the business problem, the sample used, and the mathematical and QA models. Section \ref{section:Results} presents the results, and Section \ref{section:Discussion} discusses them. Finally, Section \ref{section:Conlcusions} concludes the paper with the final conclusions and suggestions for future work.

\section{Related Work}\label{section:RelatedWork}
Syrichas and Crispin \cite{10} propose a simulated Quantum Annealing solver using Quantum Monte Carlo simulation and apply the model to large-scale benchmark datasets. They obtain optimal results by empirically manipulating the hyper-parameters of their model. This work simulates a quantum system through statistical computation, therefore, does not provide an assessment of the real system and the impact of quantum errors.

Borcinova \cite{3} describes a flow-based formulation and a hybrid
quantum annealing algorithm for solving the Vehicle Routing Problem (VRP) using
quantum annealing. The model focuses on designing directly applicable
algorithms to solve routing problems in actual companies. The author suggests that the flow base is superior to other approaches as it reduces the number of binary variables.

Borowski et al. \cite{4} introduce a hybrid algorithm, DBCAN Solver and
Solution Partitioning Solver (SPS), which uses quantum annealing to solve the
VRP and the  CVRP variant. Their experiments indicate that the
hybrid method gives promising results and can find solutions of similar
or even better quality than the tested classical algorithms. However, their results are limited to a few nodes.

Jain \cite{5} shows how to solve the  Travel Salesman Problem (TSP) problem by using an Ising
Hamiltonian-based quantum annealer and transforming it in a quadratic
unconstrained binary optimisation (QUBO) problem. They suggest that QA can only handle a minor problem (8 or fewer nodes), and even in these cases, the performance in terms of time and accuracy is subpar compared to the classical solver.

Salehi et al. \cite{6} provide a detailed analysis of the Travelling
Salesman Problem with Time Windows (TSPTW) in the context of solving it
on a quantum computer. They introduce unconstrained quadratic binary
optimisation and higher-order binary optimisation formulations of this
problem. They demonstrate the advantage of edge-based and node-based
formulation of the TSPTW problem.

Feld et al. \cite{7} investigate different quantum-classic hybrid approaches to solve the CVRP and expose the
difficulties in finding feasible solutions. They propose a
hybrid method based on a 2-Phase-Heuristic to address these limitations. After running their simulations, they concluded that the
critical step was to find an effective way to map the optimisation into
the QUBO formulation.

The analysis of this literature suggests that:

\begin{enumerate}
\def\labelenumi{\arabic{enumi}.}
  \item QA simulated on classical hardware can solve large CVRP instances and find optimal solutions by empirically evolving the model hyper-parameters.
  \item Previous studies using real quantum processors solve only small problems ($\leq$ 8 nodes), and even in this case, the results are subpar compared to the classical solver.
  \item The quality of the solution depends on the effective way of mapping the problem to a suitable formulation for the quantum solver (QUBO formulation).
\end{enumerate}

\section{Aim}\label{Aim}
This empirical work is organised around \ref{lastquestion} questions:
%
%
\begin{compactenum}[1.]
  \item \label{aim:accurracy.CVRP} What is the QA solver's accuracy for the CVRP problem over benchmark data sets?
  \item \label{aim:impact.size.accuracy} How does the size (number of nodes and routes) impact the QA solver's accuracy?
  \item \label{aim:impact.cdensity.accuracy} How does the constraint density impact the accuracy?
  \item \label{aim:impact.size.qtime} How does the problem complexity impact the time the quantum processor unit uses? \label{lastquestion}
\end{compactenum}

To answer question~\ref{aim:accurracy.CVRP}, for each problem instance,  we run 100 simulations, and we calculate the QA results accuracy using the Mean Absolute Percentage Error (MAPE) \(R\) calculated as follows:
\begin{equation}
\label{eq:22}
R_{n} = \frac{1}{n}\sum_{k = 0}^{n}{\frac{\lvert E^{k}_{QA}-E_{best} \lvert}{ E_{best} }}
\end{equation}

where:
\(E_{\text{best}}\) is the best-known solution for the instance, \(E^{k}_{\text{QA}}\) is
the QA result for the iteration \(k\) and \(R_{n}\) is the MAPE after  \(n\) iterations.

With respect to question~\ref{aim:impact.size.accuracy}, we express the size of the problem expressed as the number of binary variables in the QUBO formulation and use this measure to investigate question~\ref{aim:impact.size.qtime} and analyse how it  impacts the Quantum Process Unit (QPU) time.

Regarding question~\ref{aim:impact.cdensity.accuracy}, we express the constraint density as the tightness of the sample expressed as the total demand divided by the vehicle capacity:
\begin{equation}
\label{eq:tightnesse}
\tau = \frac{\sum_{i=0}^{n}{d_{i}}}{k \times Q}
\end{equation}

where:
\(d_{i}\) is the demand for the node \(i\), \(k\) is the number of trucks and \(Q\) is the capacity for each truck (we assume that each truck has the same capacity).

 We use \(\tau\) to measure the model complexity.
\section{Method}\label{section:Method}
\begin{figure}[htbp]
\centerline{\includegraphics{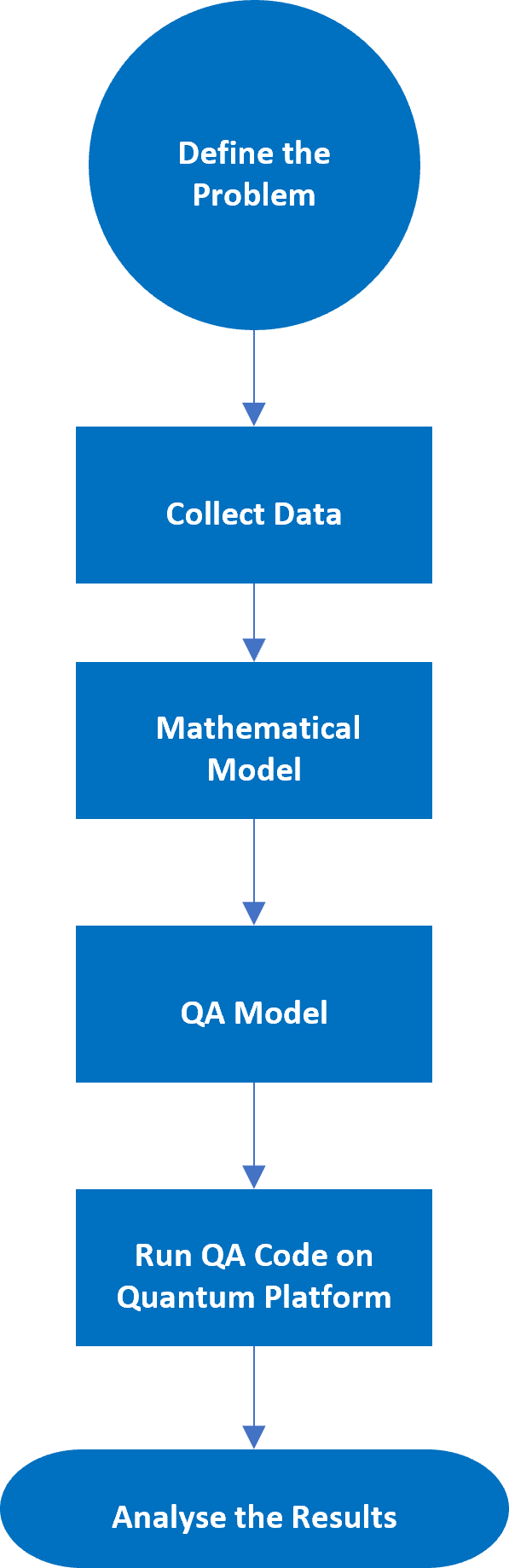}}
\caption{Quantum Annealing Optimisation Process.}
\label{figure:annealing-process}
\end{figure}

In Fig.~\ref{figure:annealing-process} we provide an overview of the process of applying quantum annealing optimisation to the selected problem.

We followed these
steps:

\begin{itemize}
\item
  Define the problem
\item
  Collect the Data
\item
Define the mathematical model 
\item
Translate the mathematical model into a QA model 
\item 
Run the QA code on the chosen platform
\item
Analyse the results
\end{itemize}

The following sections will provide details of these steps.

\subsection{The Business Problem}
The CVRP is a variation of the well-known vehicle routing problem (VRP), one of the most studied NP-hard optimisation problems, and it is usually used as a benchmark for new algorithms and computing capabilities.

In the CVRP problem, the objective is to minimise the cost of deliveries
across all customers and all routes, given that (constraints):

\begin{enumerate}
\def\labelenumi{\arabic{enumi}.}
\item
  A node is visited only once
\item
  Each vehicle can leave the depot only once
\item
  Each vehicle starts and ends its route at the depot
\item
  Each customer's demand is indivisible, and each vehicle shall not
  exceed its maximum load capacity
\item
  No route is disconnected from the depot (i.e. sub-routing elimination)
\end{enumerate}

We made the following simplifying assumptions:

\begin{enumerate}
\def\labelenumi{\arabic{enumi}.}
\item
  All vehicles have the same capacity
\item
  All vehicles have the same cost per unit distance travelled
\item
  Demands, distances between nodes, and
  delivery costs are known.
\end{enumerate}

\subsection{Sample}
%
%
We have used the data set produced by Augera et al. \cite{11}, known as the A-series. The A-series is de facto benchmark dataset to assess solutions for the CVRP.

\begin{table}[htbp]
\caption{A-CVRP Benchmark Dataset for CVRP (in bold the three instances investigated in this paper)}
\begin{center}
\begin{tabular}{|c|c|c|c|c|}
\hline
\textbf{Problem Name} & \textbf{Total Nodes} & \textbf{Vehicles} & \textbf{$\tau$} &\textbf{Best Solution} \\
\cline{1-5}
\textbf{A-n32-k5}	&	\textbf {31}	&	\textbf {5}	&\textbf {0.820}	& \textbf {784}	\\
\cline{1-5}
A-n33-k5	&	32	&	5	&0.948	& 661	\\
\cline{1-5}
A-n33-k6	&	32	&	6	&0.901	& 742	\\
\cline{1-5}
A-n34-k5	&	33	&	5	&0.920	& 778	\\
\cline{1-5}
A-n36-k5	&	35	&	5	&0.884	& 799	\\
\cline{1-5}
A-n37-k5	&	36	&	5	&0.814	& 669	\\
\cline{1-5}
A-n37-k6	&	36	&	6	&0.950	& 949	\\
\cline{1-5}
A-n38-k5	&	37	&	5	&0.967	& 730	\\
\cline{1-5}
A-n39-k5	&	38	&	5	&0.950	& 822	\\
\cline{1-5}
A-n39-k6	&	38	&	6	&0.876	& 831	\\
\cline{1-5}
A-n44-k6	&	43	&	6	&0.950	& 937	\\
\cline{1-5}
A-n45-k6	&	44	&	6	&1.050	& 944	\\
\cline{1-5}
A-n45-k7	&	44	&	7	&0.950	& 1146	\\
\cline{1-5}
A-n46-k7	&	45	&	7	&0.861	& 914	\\
\cline{1-5}
A-n48-k7	&	47	&	7	&0.892	& 1073	\\
\cline{1-5}
A-n53-k7	&	52	&	7	&0.948	& 1010	\\
\cline{1-5}
A-n54-k7	&	53	&	7	&0.955	& 1167	\\
\cline{1-5}
A-n55-k9	&	54	&	9	&0.932	& 1073	\\
\cline{1-5}
\textbf{A-n60-k9}	&	\textbf{59}	&	\textbf{9}	&\textbf{0.921}	& \textbf{1354}	\\
\cline{1-5}
A-n61-k9	&	60	&	9	&0.983	& 1034	\\
\cline{1-5}
A-n62-k8	&	61	&	8	&0.916	& 1288	\\
\cline{1-5}
A-n63-k9	&	62	&	9	&0.970	& 1616	\\
\cline{1-5}
A-n63-k10	&	62	&	10	&0.932	& 1314	\\
\cline{1-5}
A-n64-k9	&	63	&	9	&0.942	& 1401	\\
\cline{1-5}
A-n65-k9	&	64	&	9	&0.974	& 1174	\\
\cline{1-5}
A-n69-k9	&	68	&	9	&0.938	& 1159	\\
\cline{1-5}
\textbf{A-n80-k10}	&	\textbf{79}	&	\textbf{10}	&\textbf{0.948}	& \textbf{1763}	\\     
\cline{1-5}
\end{tabular}
\label{tab:dataset}
\end{center}
\end{table}

From Table \ref{tab:dataset}, the A-n32-k5 is the smallest instance, while the A-n80-k10 is the largest. The A-n60-k9 seats in the middle of these two instances. We have, therefore, intensively investigated the following three instances: 
\begin{itemize}
\item
  A-n32-k5: 32 nodes and 5 trucks.
\item
  A-n60-k9: 60 nodes and 9 trucks
\item
  A-n80-k10: 80 nodes and 10 trucks
\end{itemize}

\subsection{Mathematical Model}
Following \cite{3}, we express the objective function as follows:
\begin{equation}
\label{eq:10}
Minimize\ \sum_{r = 1}^{p}{\sum_{i = 0}^{n}{\sum_{j = 0,i \neq j}^{n}C_{\text{ij}}}}x_{\text{rij}}
\end{equation}

where \(C_{\text{ij}}\) is the distance (cost) between node \(i\) and node \(j\) and the binary variable \(x_{\text{rij}}\) is:
\begin{equation}
\label{eq:11}
x_{rij}= 
\Biggl
\{ \begin{matrix}
1\ \text{if\ truck\ r\ travels\ from\ i\ to\ j} \\
0\ \text{otherwise} \\
\end{matrix}
\end{equation}

and \(p\) is the number of trucks, \(n\) is the number of
cities (nodes) (including the depot (0)).

The following equation ensures that each node is visited only once by any truck:
\begin{equation}
\label{eq:12}
\sum_{r = 1}^{p}{\sum_{i = 0,i \neq j}^{n}{x_{\text{rij\ \ }} = 1}}\ \forall j \in \left\{ 1,..,n \right\}
\end{equation}

The following equation ensures that each vehicle visits the
depot:
\begin{equation}
\label{eq:13}
\sum_{j = 1}^{n}{x_{r0j} =  1}\ \forall r \in \left\{ 1,..,p \right\}
\end{equation}

Each vehicle has to start and end its route at the depot:
\begin{equation}
\label{eq:14}
\sum_{i = 0,i \neq j}^{n}x_{\text{rij}} = \sum_{i = 0}^{n}x_{\text{rji}}\ ,\ \forall j \in \left\{ 0,\ldots,n \right\},\text{\ \ r} \in \left\{ 1,\ldots,p \right\}
\end{equation}

The load carried by each vehicle should not exceed its capacity:
\begin{equation}
\label{eq:15}
\sum_{i = 0}^{n}{\sum_{j = 1,i \neq j}^{n}{d_{j}x_{\text{rij}} \leq Q},\ \forall r \in \left\{ 1,\ldots,p \right\}}
\end{equation}

where \(Q\) is the vehicle capacity.

The routes must be interconnected (i.e. no isolated loops): this constraint is called the "sub-routing elimination constraint" (SEC). Many SECs have been proposed.

One of the best-known is the  Dantzig, Fulkerson and Johnson (DFJ) SEC formulation given by equation (\ref{eq:16}):

\begin{equation}
\label{eq:16}
\sum_{r = 1}^{p}{\sum_{i \in S}^{n}{\sum_{j \in S,i \neq j}^{n}{x_{\text{rij\ \ }} \leq \left| S \right| - 1,\ }}}
\end{equation}

This formulation introduces  \(2^{n} + 2n - 2\) equations  and \(n(n - 1)\) ancillary variables \cite{13}.
Other authors have developed alternative SEC formulations to reduce the number of constraints and ancillary variables. For an exhaustive analysis, see \cite{17}.
One of the most used is the Miller-Tucker-Zemlin (MTZ) SECs \cite{13}.

The MTZ SEC uses an extra variable \(u_{i}\) that gets a value for each node, except for the depot. If a vehicle
drives from node \(i\) to node \(j\), the value of \(u_{j}\) has to be bigger than
the value of \(u_{i}\).

The mathematical formulation of the MTZ SEC is as follows:
\begin{equation}
\label{eq:17}
u_{j} - u_{i}\  \geq \ q_{j} - Q\left( 1 - x_{\text{ijk}} \right)\\
\forall i,j\  \in V\backslash\left\{ 1 \right\}\ i \neq j
\end{equation}

\begin{equation}
\label{eq:18}
q_{i} \leq u_{i} \leq Q\\
\forall i\  \in V\backslash\left\{ 1 \right\}
\end{equation}

If vehicle \(k\) drives from node \(i\) to node \( j\) ,
\(x_{\text{ijk}} = 0,\) and constraint \ref{eq:17} can be written to
\(u_{j} \geq u_{i} + q_{j}\). This ensures that the value of \(u_{j}\)
is at least \(q_{j}\) more than \(u_{i}\). So, the value of \(u_{j}\) is
greater than the value of \(u_{i}\).

The MTZ SEC introduces \(n^{2} - n + 2\) constraints, \(n(n - 1)\) 0--1 variables, and
\((n - 1)\) continuous variables.

\begin{table}[htbp]
\caption{Size comparisons for DFJ and MTZ sub-tour elimination formulations.}
\begin{center}
\begin{tabular}{|c|c|c|c|}
\hline
\textbf{Formulation} & \textbf{Variables} & \textbf{Constraints} & \textbf{Constraints for n=20} \\
\cline{1-4}
\hline
DFJ & O(n\textsuperscript{2}) & O(2\textsuperscript{n}) & 2\textsuperscript{n}=1,048,576\\
\hline
MTZ & O(n\textsuperscript{2}) & O(n\textsuperscript{2}) & n\textsuperscript{2}=400\\
\hline    
\end{tabular}
\label{tab:2}
\end{center}
\end{table}

Table \ref{tab:2} compares the MTZ and DFJ SEC formulations and shows that MTZ's approach adds a polynomial number of constraints,
while DFJ's approach introduces an exponential number of constraints \cite{13}. For this reason, the MTZ SEC formulation has been adopted.
\subsection{QA Model}
Any given NP-hard problem instance can be translated into an Ising Model (IM) instance with no more than a polynomial expansion in problem size. Any Ising Model instance of \(n\) variables can be minor-embedded onto a quantum annealing graph using \(O(n^{2})\) qubits in the worst case. In essence, quantum annealing processors are designed to solve objective functions  expressed as Ising Model (IM). 
It is easy to prove that the Ising formulation is equivalent to a Quadratic Unconstrained Binary Optimisation (QUBO) formulation, given a simple variable substitution \{-1,1\} (IM) into \{0,1\} (QUBO) \cite{8}.
For this association with the Ising problem, the QUBO model has emerged as an underpinning of quantum annealing and lies at the heart of experimentation with quantum computers.

This  section briefly introduces the generic QUBO formulation and how it is adapted to include constraints.

Let's consider the optimisation problem:
\begin{equation}
\label{eq:1}
\text{Minimise} \ y = - 3x_{1} - 5x_{2} + 2x_{1}x_{2}
\end{equation}
where the variables $x_{i}$ are binary {0,1}.

The above function is quadratic in binary variables with a linear part
-3$x_{1}$ -5$x_{2}$ and a quadratic part $2x_{1}x_{2}$~.~

As $x_{i}$ are binary, $x_{i}$=$x_{i}^{2}$, the linear part can be
written as -3$x_{1}^{2}$ -5$x_{2}^{2}$.

We can rewrite the model using matrices. Let's consider the optimisation problem:

\begin{equation}
\label{eq:2}
\text{Minimise}\ y = \left( x_{1}\ x_{2} \right)\begin{bmatrix}
 - 3 & 2 \\
0 & - 5 \\
\end{bmatrix}\begin{bmatrix}
x_{1} \\
x_{2} \\
\end{bmatrix}
\end{equation}
The matrix notation of this can be written as
$Minimise\ y = \ \mathbf{x}^{T}Q\mathbf{x}$, where $\mathbf{x}$ is a column vector of binary variables.

The coefficients of the original linear terms appear on the main
diagonal of the $Q$ matrix.

From a mathematical point of view, the problem's constraints are
inequalities that the solution must respect. By default, the QUBO
formulation does not allow constraints. To include constraints, we need
to rewrite them as quadratic equations and introduce them as penalties
that influence the value of the objective function.

Penalties are formulated to have zero value for feasible solutions and a
positive amount for invalid solutions. For inequalities, slack variables
transform them into equalities. Each penalty should be multiplied by a
positive constant to have a comparable magnitude with the objective
function; we denote these constants with $P$ (called the Lagrange
multiplier).

For example, assuming that one constraint is:
\begin{equation}
\label{eq:3}
x_{1} + x_{2} = 1
\end{equation}
Such constraints can be formulated as follows:
\begin{equation}
\label{eq:4}
P*\left( x_{1} + x_{2} - 1 \right)^{2}
\end{equation}
The objective function becomes:
\begin{equation}
\label{eq:5}
\text{Minimise}\ y = - 3x_{1} - 5x_{2} + 2x_{1}x_{2} + 
P*\left( x_{1} + x_{2} - 1 \right)^{2}
\end{equation}

Expanding the constraint and considering $x_{i}$=$x_{i}^{2}$
\begin{equation}
\label{eq:6}
Q = \begin{bmatrix}
 - 3 & 2 \\
0 & - 5 \\
\end{bmatrix} + P\begin{bmatrix}
 - 1 & 2 \\
0 & - 1 \\
\end{bmatrix}
\end{equation}

Constraints expressed as inequalities need slack variables. For example:
\begin{equation}
\label{eq:7}
2x_{1} + {3\ x}_{2}\  \leq C
\end{equation}
where \(C\) is a positive quantity.
Can be expressed as:
\begin{equation}
\label{eq:8}
P*\left( 2\ x_{1} + 3x_{2} - C + \ \sum_{k = 0}^{r_{u}}{2^{k}v_{k}} \right)^{2}
\end{equation}
Where $v_{k} \in \left\lbrack 0,1 \right\rbrack$ is the slack variable
and $r_{\text{u\ }}$is such that:
\begin{equation}
\label{eq:9}
C \approx \sum_{k = 0}^{r_{u}}{2^{k}v_{k}}
\end{equation}

From this analysis, we conclude that:

1)  The number of model parameters (also called the Lagrange
    multipliers) increases as the number of constraints increases

2)  The introduction of inequalities introduced additional slack
    variables that consume physical resources on the quantum platform

As we don't know the value of the Lagrange variable before running the
model, we have to run multiple simulations to find the suitable range
for these coefficients empirically.
\subsection{Run on Quantum Platform} \label{quantumsolution}
D-Wave has developed and commercialised quantum annealing processor units designed to solve Ising Models. Their hardware can be programmed via low-level Quantum Machine Instruction or a standard set of Internet APIs based on RESTFul service. Client libraries are available in many programming languages, including Python SDK \cite{1}.
We have used the Python SDK to access the system as a cloud resource over the Internet. In our simulations, we have used the D-Wave Leap hybrid quantum solver: this service can be used to submit an arbitrary quadratic model. In November 2022, D-Wave introduced a new model called Constrained Quadratic Model (CQM) \cite{1}. CQM can be used for problems with binary and integer variables and one or more constraints. In contrast to previous hybrid solvers, which required that  any problem constraints be modelled as a penalty model in the objective function, the CQM solver natively supports equality and inequality constraints through symbolic maths. As the CQM is a new solver,  no empirical studies have analysed its performance to the best of our knowledge.
\section{Results}\label{section:Results}
\begin{figure}[htbp]
     \centering
     \begin{subfigure}[b]{0.5\textwidth}
         \centering
         \includegraphics[width=\textwidth]{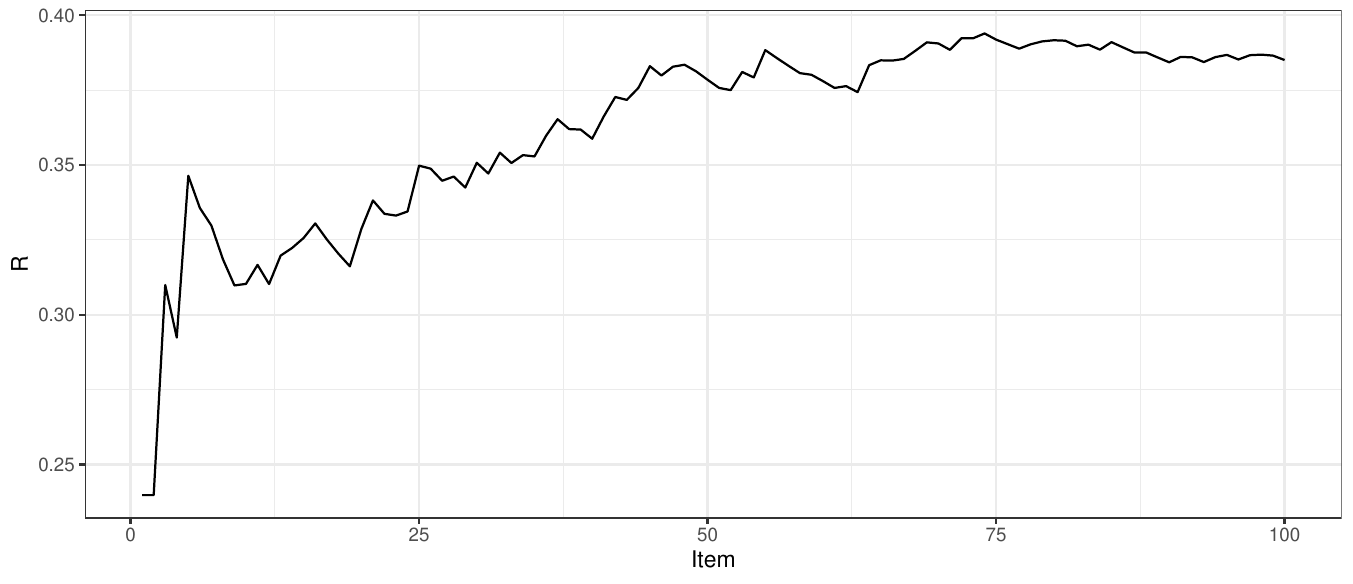}
         \caption{A-n32-k5}
         \label{fig:mape-32-5}
     \end{subfigure}
     \hfill
     \begin{subfigure}[b]{0.5\textwidth}
         \centering
         \includegraphics[width=\textwidth]{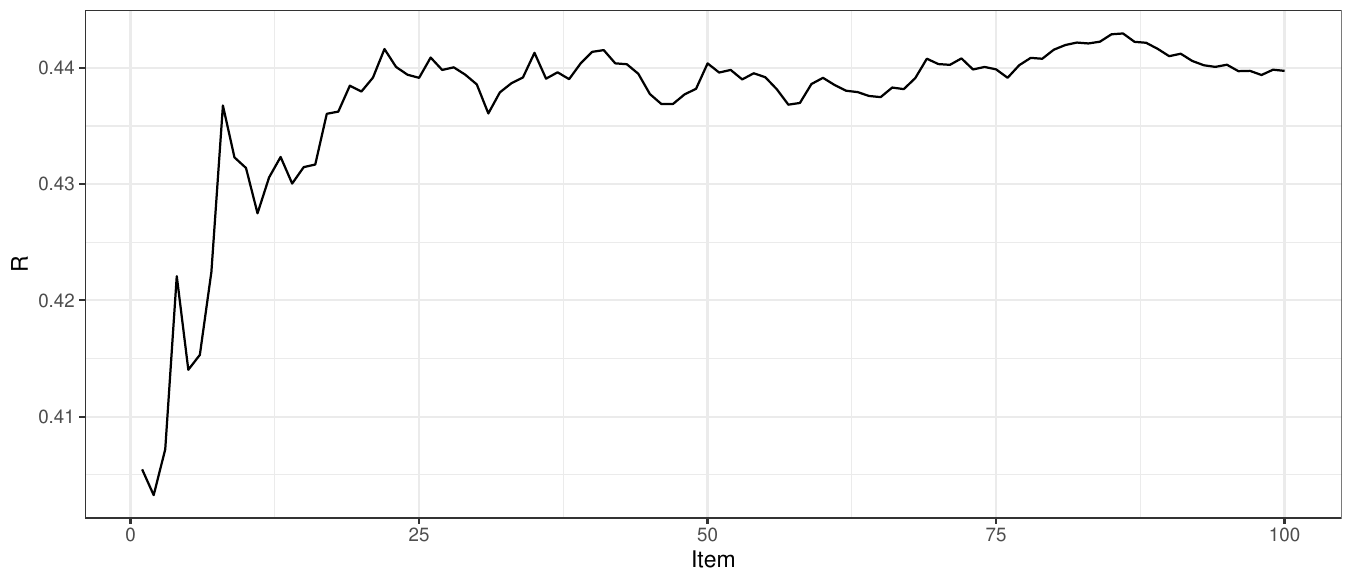}
         \caption{A-n60-k9}
         \label{fig:mape-60-9}
     \end{subfigure}
     \hfill
       \begin{subfigure}[b]{0.5\textwidth}
         \centering
         \includegraphics[width=\textwidth]{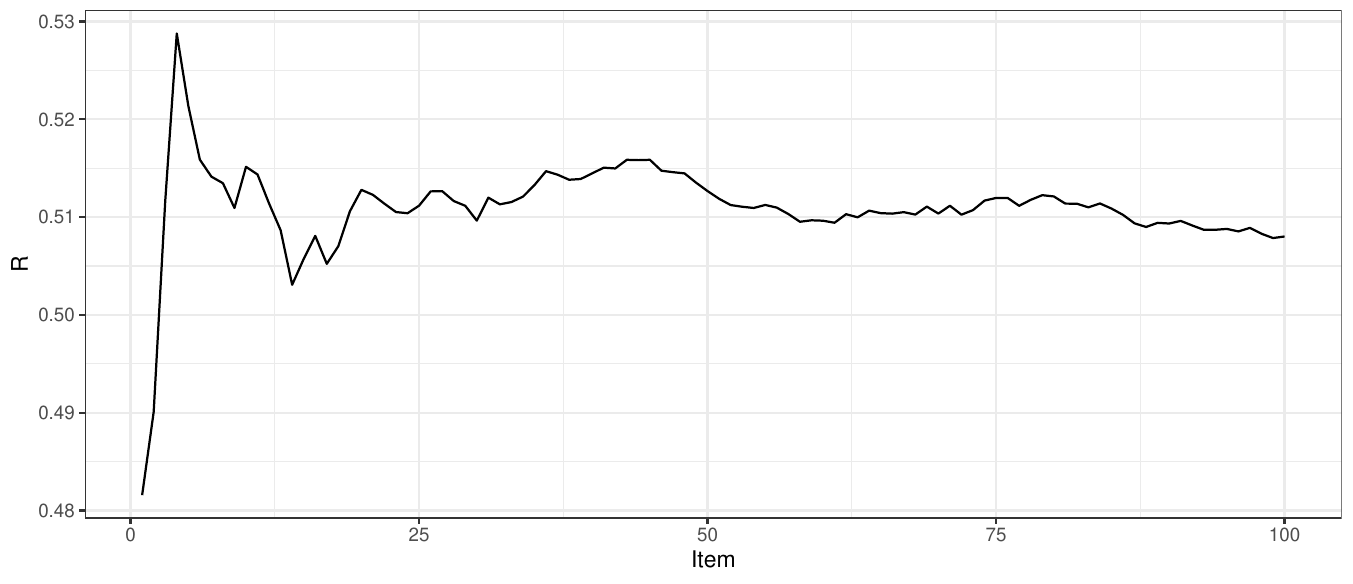}
         \caption{A-n80-k10}
         \label{fig:mape-80-10}
     \end{subfigure}
     \hfill
    
        \caption{MAPE}
        \label{fig:mape}
\end{figure}

\begin{figure}[htbp]
     \centering
     \begin{subfigure}[b]{0.5\textwidth}
         \centering
         \includegraphics[width=\textwidth]{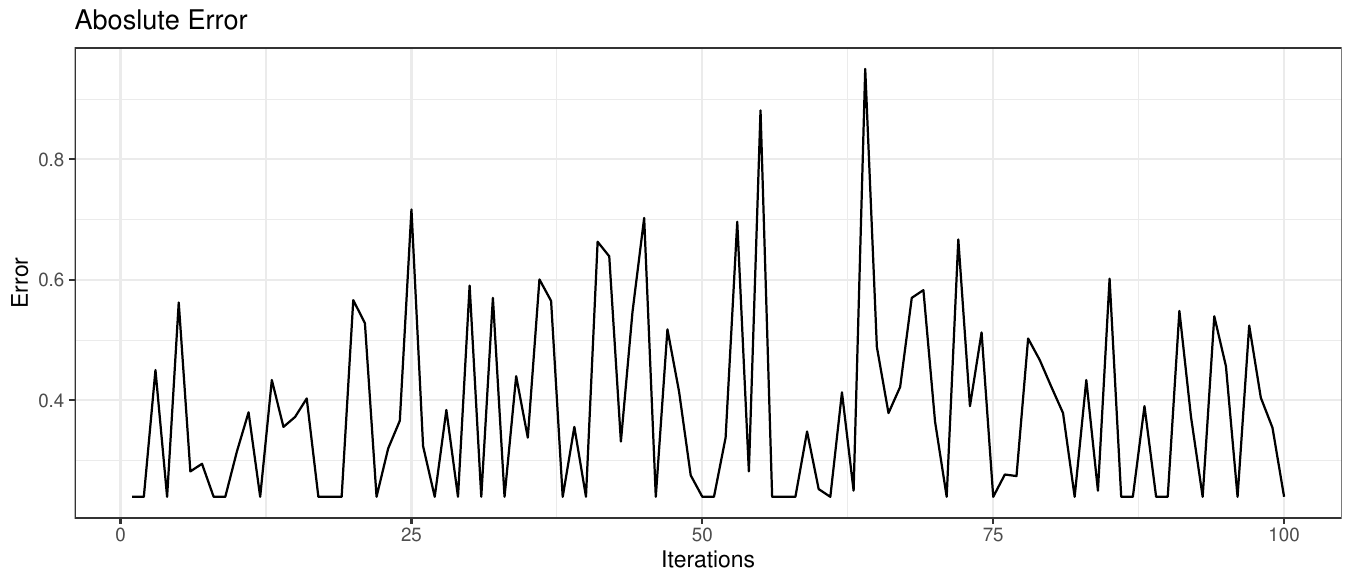}
         \caption{A-n32-k5}
         \label{fig:aerr-32-5}
     \end{subfigure}
     \hfill
     \begin{subfigure}[b]{0.5\textwidth}
         \centering
         \includegraphics[width=\textwidth]{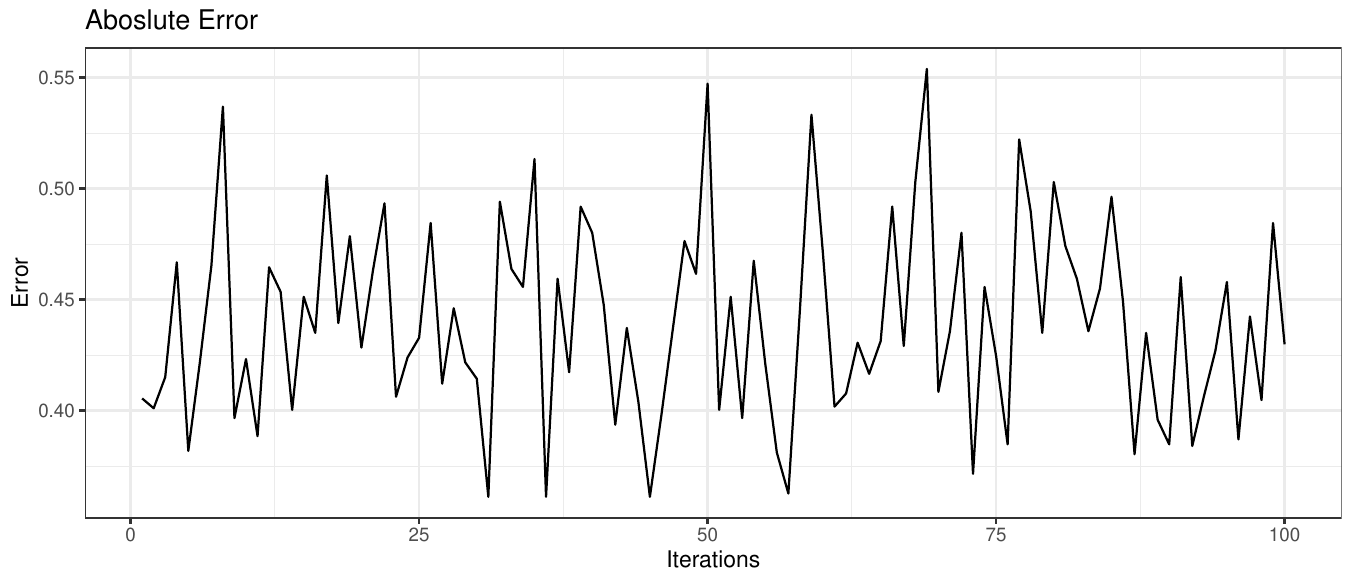}
         \caption{A-n60-k9}
         \label{fig:aerr-60-9}
     \end{subfigure}
     \hfill
       \begin{subfigure}[b]{0.5\textwidth}
         \centering
         \includegraphics[width=\textwidth]{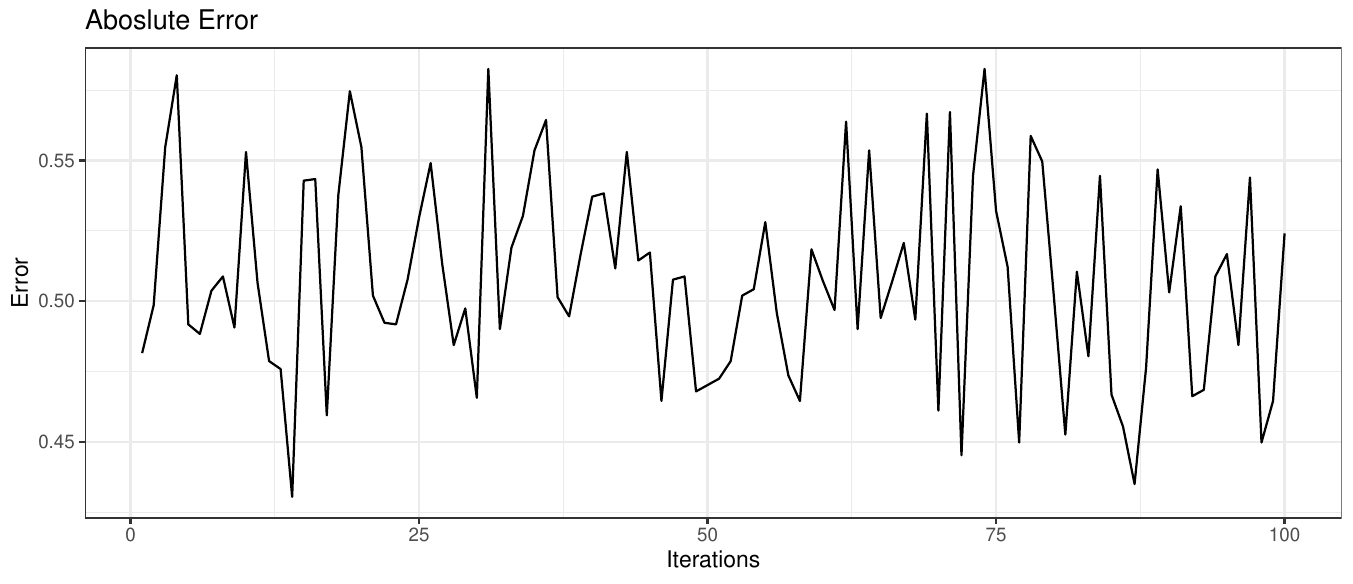}
         \caption{A-n80-k10}
         \label{fig:aerr-80-10}
     \end{subfigure}
     \hfill
    
        \caption{Absolute Error}
        \label{fig:absolute error}
\end{figure}

We consider solution statistics aggregated over 100 distinct run per instance, and we calculate the quantum solution accuracy as MAPE index as described in section \ref{Aim}. Fig. ~\ref{fig:mape} shows the MAPE for the three instances.
 We evaluate the absolute error, defined as:

\begin{equation}
\label{eq:absolute_difference}
AE_{n} = \frac{\lvert E^{n}_{QA}-E_{best} \lvert}{ E_{best} }
\end{equation}

where:
\(E_{\text{best}}\) is the best-known cost for the instance reported in the CVRP library http://vrp.atd--lab.inf.puc--rio.brlib, \(E^{n}_{\text{QA}}\) is
the QA result for the iteration \(n\). Fig. ~\ref{fig:absolute error} shows the \(AE_{n}\) for the three instances.

Table \ref{tab:5} summarises the QPU time in $\mu$s and shows that the average execution time on the QPU is similar for the three problems. For each problem, the standard deviation for the QPU time is small: the QPU time ranges from $\mu$s 32 for the A-n32-k5 to $\mu$s 48 for the A-n80-k10 sample. 

\begin{table}[htbp]
\caption{QPU time.}
\begin{center}
\begin{tabular}{|c|c|c|c|c|}
\hline
\textbf{Time ($\mu$s)} &  \textbf{A-n32-k5} & \textbf{A-n60-k9} & \textbf{A-n80-k10} \\
\cline{1-4}
Avg QPU time  & 32 & 48 & 48\\
\cline{1-4}
Std. Dev. QPU time & 0.8 & 0.6 & 0.5\\
 \cline{1-4}  
\end{tabular}
\label{tab:5}
\end{center}
\end{table}

\begin{table}[htbp]
\caption{Model Results.}
\begin{center}
\begin{tabular}{|c|c|c|c|c|}
\hline
 &  \textbf{A-n32-k5} & \textbf{A-n60-k9} & \textbf{A-n80-k10} \\
\cline{1-4}
Best Known  Solution& 784 & 1354 & 1763\\
\cline{1-4}
QA Best Solution & 972 & 1843 & 2522 \\
\cline{1-4}
QA Worst Solution & 1529 & 2104 & 2790\\
\cline{1-4}
QA Average & 1085 & 1949 & 2660\\
\cline{1-4}
QA Std. Dev. & 124 & 58 & 63 \\
\cline{1-4}
MAPE (100)  & 0.36 & 0.43 & 0.51 \\
\cline{1-4}
Min QPU time ($\mu$s) & 30 & 45 & 46\\ 
\cline{1-4}  
\end{tabular}
\label{tab:4}
\end{center}
\end{table}

\begin{table}[htbp]
\caption{Models Density.}
\begin{center}
\begin{tabular}{|c|c|c|c|c|}
\hline
 \textbf{Input} & \begin{tabular}{clr} \#CQM \\Variables\end{tabular}  & 
  \begin{tabular}{clr} \#CQM \\Constraints\end{tabular}  &
      \begin{tabular}{clr} \#CQM \\Biases\end{tabular} &
      \textbf{$\tau$} \\
\cline{1-5}
 A-n32-k5 & 5115 & 4851 & 38750 & 0.820\\
\cline{1-5}
 A-n60-k9 & 32391 & 31415 & 251694 & 0.921\\
 \cline{1-5}  
  A-n80-k10 & 63990 & 62519 & 500860 & 0.941\\
\cline{1-5}
\end{tabular}
\label{tab:3}
\end{center}
\end{table}

%
%

\begin{figure}[htbp]
     \centering
     \begin{subfigure}[b]{0.5\textwidth}
         \centering
         \includegraphics[width=\textwidth]{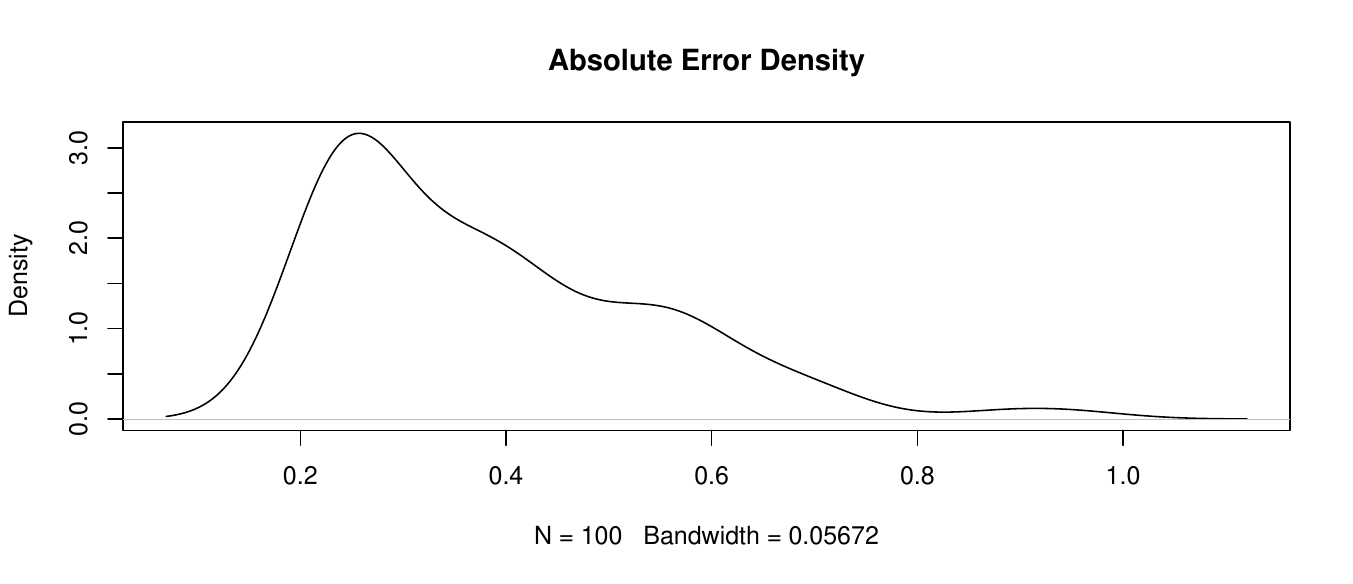}
         \caption{A-n32-k5}
         \label{fig:derr-32-5}
     \end{subfigure}
     \hfill
     \begin{subfigure}[b]{0.5\textwidth}
         \centering
         \includegraphics[width=\textwidth]{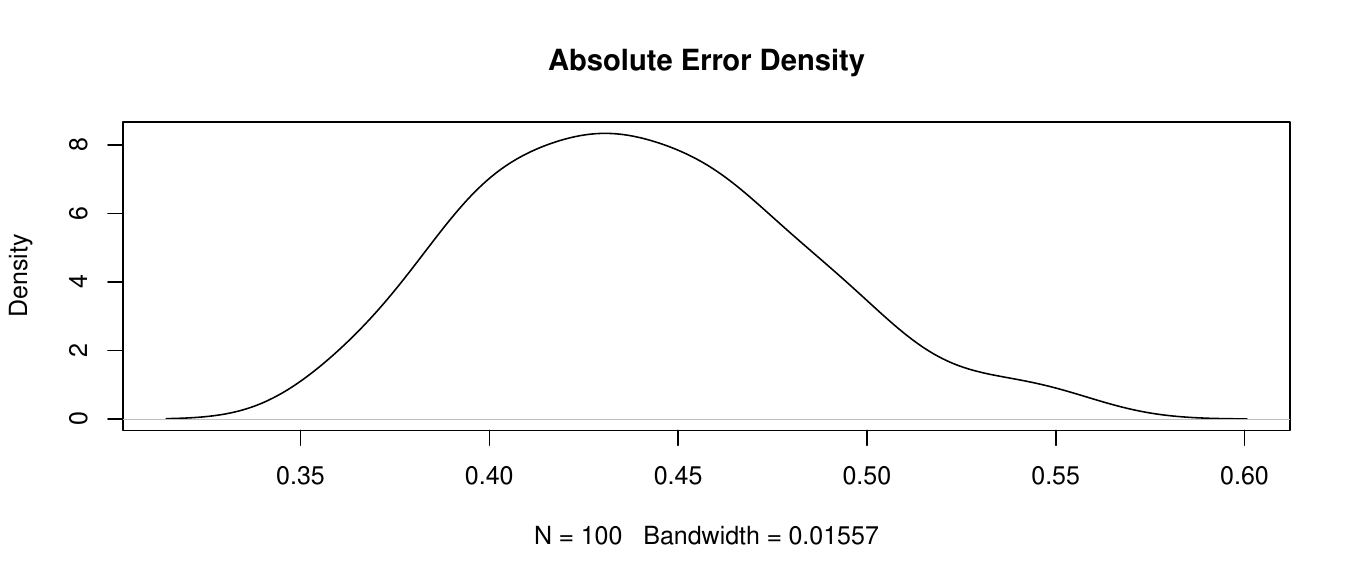}
         \caption{A-n60-k9}
         \label{fig:derr-60-9}
     \end{subfigure}
     \hfill
       \begin{subfigure}[b]{0.5\textwidth}
         \centering
         \includegraphics[width=\textwidth]{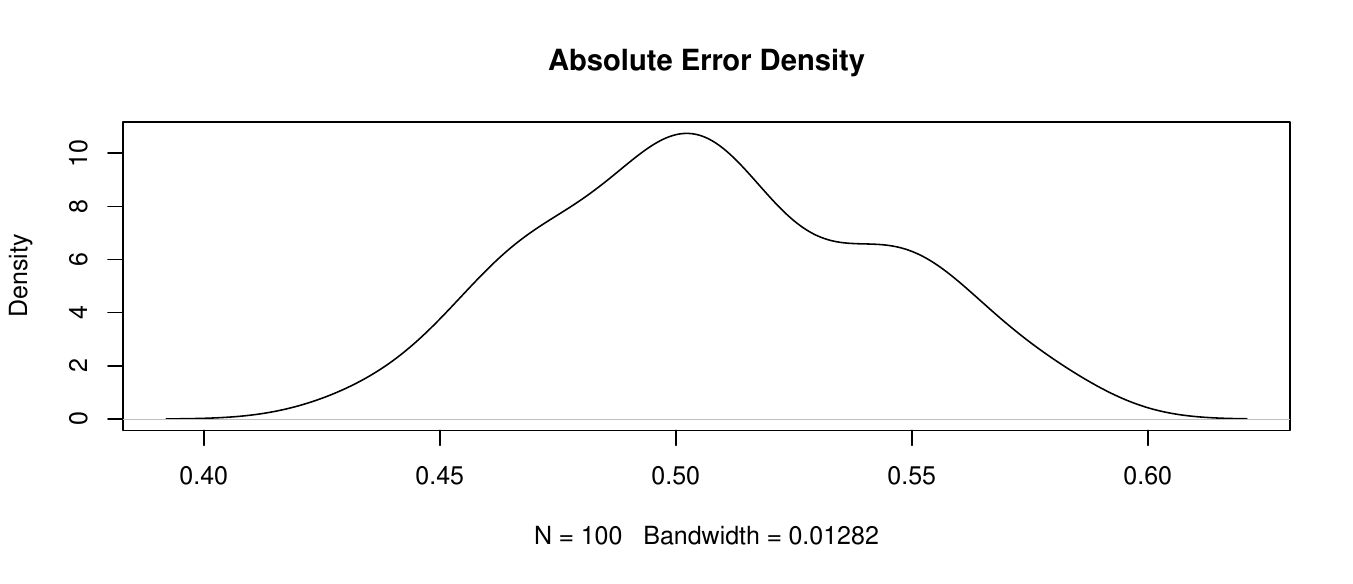}
         \caption{A-n80-k10}
         \label{fig:derr-80-10}
     \end{subfigure}
     \hfill
    
        \caption{Error Density}
        \label{fig:error density}
\end{figure}

%
%
\section{Discussion}\label{section:Discussion}
Although NP-hard problems expressed in QUBO formulation can be solved using QA, it is challenging to develop an efficient model because the QA solver's performance depends critically on the problem's complexity and size. The QUBO model generally becomes cumbersome as the number of variables and constraints grows.

For the A-n32-k5 problem, the MAPE increases during the first 75 simulations (Fig. \ref{fig:mape-32-5}) when it stabilises around 0.36. The MAPE for the A-n60-k9 (Fig. \ref{fig:mape-60-9}) stabilises earlier (after about 50 iterations) and has a value of 0.43. This is the same for the A-n80-k10 (Fig. \ref{fig:mape-80-10}), with a value of 0.51 after 50 runs.
Table \ref{tab:4} also shows that the standard deviation of the QA solutions decreases as the problem size increases.
The absolute error densities error (Fig. \ref{fig:error density}) show that for the smaller problem (A-n32-k5), the solver is more likely to find the best solution (closer to the optimal) and shows a Pareto distribution, while as the size of the model increases, the results tend to concentrate around the average exhibiting a Gaussian distribution.
The size of the problem for the range considered (from $\approx$ 5,000 to $\approx$ 63,000 binary variables) does not impact the QPU time, and the variation between simulations is small.

\subsection{Limitations and Future Work}
Although the new D-Wave CQM solver removes the need for the manual definition of slack variables while allowing the modelling of complex quadratic constraints (e.g., the DTM sub-routing approach defines constraints that are not linear), it remains a black box approach, and the user has less control on the hyperparameters to control the execution.
It is worth noting that the QA does not guarantee the best solution. The solution's quality deteriorates as the problem's complexity increases. A way to address this limitation for future work is to split the problem into two phases, a clustering phase and a TSP solution for each cluster.

\section{Conclusions}\label{section:Conlcusions}

In this paper, we used the D-Wave CQM solver to run multiple simulations for the CVRP problem and compared the results against known optimal solutions for three datasets.
The CQM solver offers some benefits as it allows for expressing objective functions using a combination of different types of variables. It also helps express constraints using intuitive symbolic algebraic notation.
The direct QUBO formulation  could not
find any feasible solution as the number of nodes moved above 15 nodes and 3
trucks, this is consistent with other studies. As the constraints add penalty terms to the Hamiltonian of the problem, the model consumes increasing resources while also restricting the dynamic range of the interactions between nodes.
We conducted extensive simulations for each problem (more than 100 per problem) and calculated the aggregated MAPE and time to execute. The MAPE index estimates how good the actual quantum solution is
compared to the "optimal" solution. 
Our results show that we must be careful about formulating the problem and especially the constraints equations: the formulation with the least number of constraints has to be preferred to other models.

\vspace{12pt}

\end{document}